\documentclass[prd, 10pt,aps,nofootinbib, floatfix, notitlepage,twocolumn]{revtex4-1}
\usepackage{amsmath,amsfonts,amssymb}
\usepackage{mathrsfs}
\usepackage{graphicx}
\usepackage{subfigure}
\usepackage[english]{babel} 
\usepackage{url} 
\usepackage{color}
\usepackage[utf8]{inputenc} 
\usepackage{float}
\usepackage[colorlinks=true,citecolor=blue,urlcolor=blue]{hyperref}
\usepackage{booktabs}
\usepackage{subfigure}

\usepackage[T1]{fontenc}

\begin{document}


\title{Equality scale-based and sound horizon-based analysis of the Hubble tension}


\author{Zhihuan Zhou}
\email{11702005@mail.dlut.edu.cn}
\altaffiliation[Also at ]{}
\author{Yuhao Mu}
\author{Gang Liu}
\author{Lixin Xu}%
\email{lxxu@dlut.edu.cn}
\affiliation{Institute of Theoretical Physics \\
	School of Physics \\
	Dalian University of Technology \\
	Dalian 116024, People's Republic of China}%


\author{Jianbo Lu}
\affiliation{
	Department of Physics\\
	Liaoning Normal University \\
	Dalian 116029, People's Republic of China
}%


\date{\today}

\begin{abstract}
The Hubble horizon at matter-radiation equality ($k^{-1}_{\rm{eq}}$) and the sound horizon at the last scattering surface ($r_s(z_*)$) provides interesting consistency check for the $\Lambda$CDM model and its extensions.
It is well known that the reduction of $r_s$ can be compensated by the increase of $H_0$, while the same is true for the standard rulers $k_{\rm{eq}}$.
Adding extra radiational component to the early universe can reduce $k_{\rm{eq}}$. The addition of early dark energy (EDE), however, tends to increase $k_{\rm{eq}}$. We perform $k_{\rm{eq}}$- and $r_s$-based analyses in both the EDE model and the Wess-Zumino Dark Radiation (WZDR) model. In the latter case we find $\Delta H_0 = 0.4$ between the $r_s$- and $k_{\rm{eq}}$-based datasets, while in the former case we find $\Delta H_0 = 1.2$. This result suggests that the dark radiation scenario is more consistent in the fit of the two standard rulers ($k_{\rm{eq}}$ and $r_s$). As a forecast analyses, we fit the two models with a mock $k_{\rm{eq}}$ prior derived from \emph{Planck} best-fit $\Lambda$CDM model. Compared with the best-fit $H_0$ in baseline $\Lambda$CDM model, we find $\Delta H_0 = 1.1$ for WZDR model and $\Delta H_0 = - 2.4$ for EDE model.

\end{abstract}


\maketitle

\section{Introduction}\label{sec:intro}
The Cepheid-calibrated supernovae Ia (SnIa) ($z\lesssim 1.6$) \cite{Riess:2020fzl,Riess:2021jrx}, the acoustic scale extracted from CMB ($z\sim 1100$) and the BAO feature measured in LSS surveys \cite{Abbott:2017wau,Alam:2016hwk,Hildebrandt:2018yau,Hikage:2018qbn} ($z\sim 0.4$) are amongst the most precise cosmic distance anchors.
Mismatches between these anchors give rise to the well-known $H_0$ discrepancy \cite{DiValentino:2021izs, Perivolaropoulos:2021jda, Abdalla:2022yfr}.
Specifically, the measured $H_0$ from local distance ladder\footnote{The local distance ladder depends on the choice of local calibrator, i.e., the Cepheid-calibrated SnIa from SH0ES found $H_0 = 73.04\pm 1.04 $\cite{Riess:2021jrx}, whilst $H_0 = 69.8\pm1.9 $\cite{Freedman:2021ahq} when the local distance ladder is calibrated by the tip of the red giant branch (TRGB) methods.} are significantly higher than that inferred from the CMB \cite{Planck:2018vyg, Chen:2018dbv} and BAO \cite{Alam:2016hwk,Gil-Marin:2018cgo,Bautista:2017zgn}, and both of the latter two relies on the determination of the sound horizon scale at either the last scattering surface ($r_s(z_*)$) or the drag epoch ($r_s(z_d)$).

Direct solutions to the $H_0$ discrepancy from a theoretical point of view\footnote{It is possible that the $H_0$ tension is caused by unknown systematics. Many recent works are focusing on this issue \cite{Mortsell:2021nzg,Perivolaropoulos:2022vql,Perivolaropoulos:2021bds}.} can roughly be categorized as as the late-time $(z\lesssim 2)$ \cite{Benevento:2020fev, Alestas:2021luu, Keeley:2019esp, Zhou:2021xov, Alestas:2021nmi,Bassett:2002qu,Li:2019yem, Hernandez-Almada:2020uyr, Efstathiou:2021ocp,Perivolaropoulos:2021bds,Vagnozzi:2021gjh,Cai:2022dkh} and the early-time ($z\gtrsim 1100$) deformations \cite{Bernal:2016gxb,Poulin:2018cxd,Smith:2019ihp, Moss:2021obd, Berghaus:2019cls, Hill:2018lfx,Murgia:2020ryi}, as well as other approaches \cite{Carneiro:2018xwq,Feng:2021ipq,Alestas:2020zol,Marra:2021fvf,Pandey:2019plg,Becker:2020hzj,Velten:2021cqj,Harvey:2015hha,Krall:2017xcw,Buen-Abad:2017gxg,Pan:2018zha,Escudero:2018thh}.
The late solutions include a phantom transition of DE which raises the CMB predicted $H_0$ value while preserving the overall shape of the CMB spectra (including $r_s$) \cite{Li:2019yem,Hernandez-Almada:2020uyr,Zhou:2021xov}, however, such modifications would violate the BAO predictions.
To shift both the CMB and BAO anchors at the same time, one may consider reducing the sound scale at last scattering, $r_s(z_*)$ by adding extra component, e.g., early dark energy \cite{Poulin:2018cxd,Hill:2018lfx,Murgia:2020ryi,Hill:2020osr,Poulin:2018dzj,Das:2020wfe,Smith:2022hwi} and dark radiation \cite{Aloni:2021eaq, Bernal:2016gxb, Seto:2021xua,Dvorkin:2014lea,Aboubrahim:2022gjb,Steigman:2013yua,Ghosh:2021axu,Vagnozzi:2019ezj}.  
In this case, the $H_0$ tension can still not be fully resolved \cite{Jedamzik:2020zmd}. One reason is that the additional component, e.g., early dark energy (EDE), would bring in new tension with the density fluctuation amplitude, $\sigma_{8}$ \cite{Hill:2020osr}.
Another is that given a shift in $r_s$, the shift in $H_0$ needed to match CMB will be different from the one needed to match the LSS \cite{DAmico:2020ods}. 

Historically, much of the focus has been put on $r_s$. To shed light on the $H_0$ discrepancy, additional standard rulers, especially those in the early universe, are urgently needed. In this work, we focus on the standard ruler $k^{-1}_{\rm{eq}}$, i.e., the Hubble horizon at matter-radiation equality, which can be derived from the turnover scale of the Full Shape (FS) power spectrum. The isolation of $k_{\rm{eq}}$ based information in Full shape (FS) galaxy power spectrum \cite{Philcox:2020xbv,Farren:2021grl} (or CMB lensing power spectrum \cite{Baxter:2020qlr}) require $r_s$-marginalization operations, which can be achieved either by excluding the prior on the baryon density $\omega_b$ \cite{Philcox:2020xbv} or removing the BAO features in the matter power spectrum via rescaling of the BAO wiggles \cite{Farren:2021grl}. It is shown in Ref. \cite{Philcox:2022sgj} that more information can be extracted from the features FS power spectrum by adopting the latter approach. When combined with CMB lensing and $\Omega_m$ prior from the recent Pantheon+ analysis \cite{Brout:2022vxf}, the standard ruler $k_{\rm{eq}}$ is able to deliver precise cosmological constraints. Especially, $k_{\rm{eq}}$ probes the features of new physics around the equality $a_{\rm{eq}}$, which is otherwise difficult to detect.

The additional EDE component which peaked around the equality scale ($\log_{10} z_c \sim 3.5$) would increase the expansion rate $H_{\rm{eq}}\equiv H(a_{\rm{eq}})$ as well as the equality scale $k_{\rm{eq}}\equiv a_{\rm{eq}}H_{\rm{eq}}$, while reducing the sound horizon scale $r_s(z_*)$. 
In mock EDE cosmology\footnote{Euclid like mock data are generated assuming EDE model. See Ref. \cite{Farren:2021grl} for detailed discussions.}, one would notice a significant peak shift in $H_0$ measured from $k_{\rm{eq}}$- to $r_s$- based analysis if $\Lambda$CDM model is assumed \cite{Farren:2021grl}. This method provides a novel consistency check of the $\Lambda$CDM cosmology.
On the other hand, the $H_0$ value measured by the two rulers should be consistent as long as the assumed model is correct.
Thus, it is of particular interest that can the modifications to the early universe shift both of the rulers ($k_{\rm{eq}}$, $r_s$) so that the inferred $H_0$ from each of the rulers be consistent with late-time measurements? 
Note that the redshift of equality $z_{\rm{eq}}$ is sensitive to the change in the radiational (matter) component. An addition of radiation component before the equality will reduce $z_{\rm{eq}}$ and thus shift $k_{\rm{eq}}$ to smaller value, 
which is in contrast with the shift brought by the extra DE component.
Given $k_{\rm{eq}}$ (measured in unit of $h\rm{Mpc}^{-1}$) and a probe of $\Omega_m$, one can solve for the Hubble constant \cite{Philcox:2020xbv}.
We show in Sec. \ref{sec:impact} that the reduction of $k_{\rm{eq}}$ will lead to a higher inferred value of $H_0$ compared with the best-fit of the $\Lambda$CDM model. 

One example of adding the radiational component is the recently investigated Wess-Zumino Dark Radiation (WZDR) model. The cosmological features of the WZDR model can be effectively described by a step transition of the effective number of neutrinos, $N_{\rm{eff}}$, when the interacting dark radiation density increases as the mediator deposits its entropy into the lighter species. 
Compared with the reference self-interacting dark radiation (SIDR) model, the transition of radiation component in the WZDR model will shift the position of the high-$\ell$ modes, which enter the horizon before the step.
It is shown in Ref. \cite{Aloni:2021eaq} that including such "stepped" dark radiation does significantly improve the combined fit to CMB, BAO, and SH0ES data while not degrading the fit to the CMB and BAO.

The outline of this paper is as follows: in Sec. \ref{sec:models} we take a brief review of the basic equations associated with the WZDR model. 
in Sec. \ref{sec:impact}, we show how EDE and WZDR model affects the standard ruler $k_{\rm{eq}}$. 
in Sec. \ref{sec:datasets}, we describe our numerical implementation of both models and the datasets used in our analysis. 
The numerical results of the Markov chain Monte Carlo (MCMC) analysis are presented in Sec. \ref{sec:results}. The discussion and conclusions are presented in Sec. \ref{sec:conclusions}.\\
\begin{figure}
	\centering
	\includegraphics[scale=0.38]{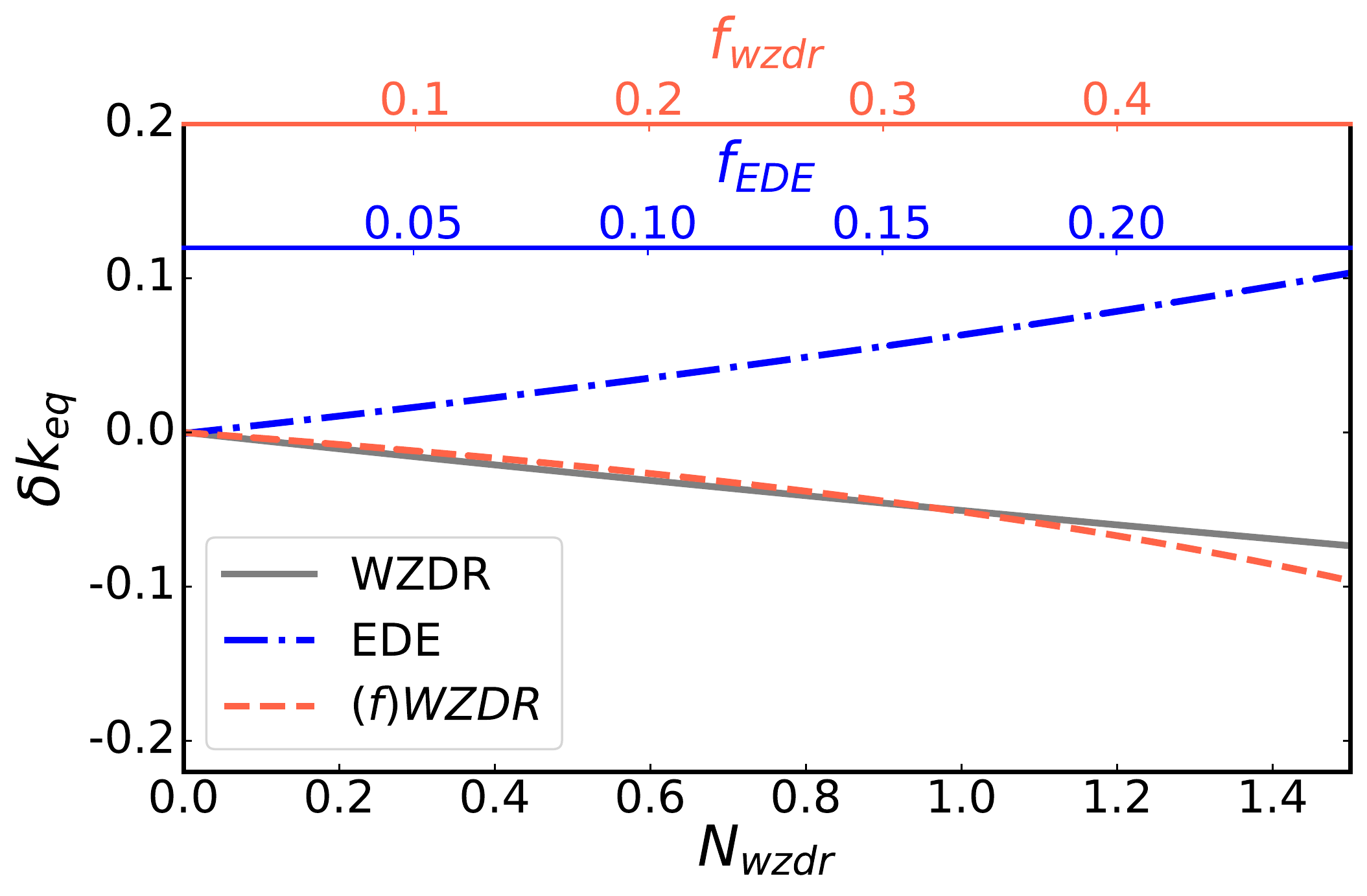}
	\includegraphics[scale=0.38]{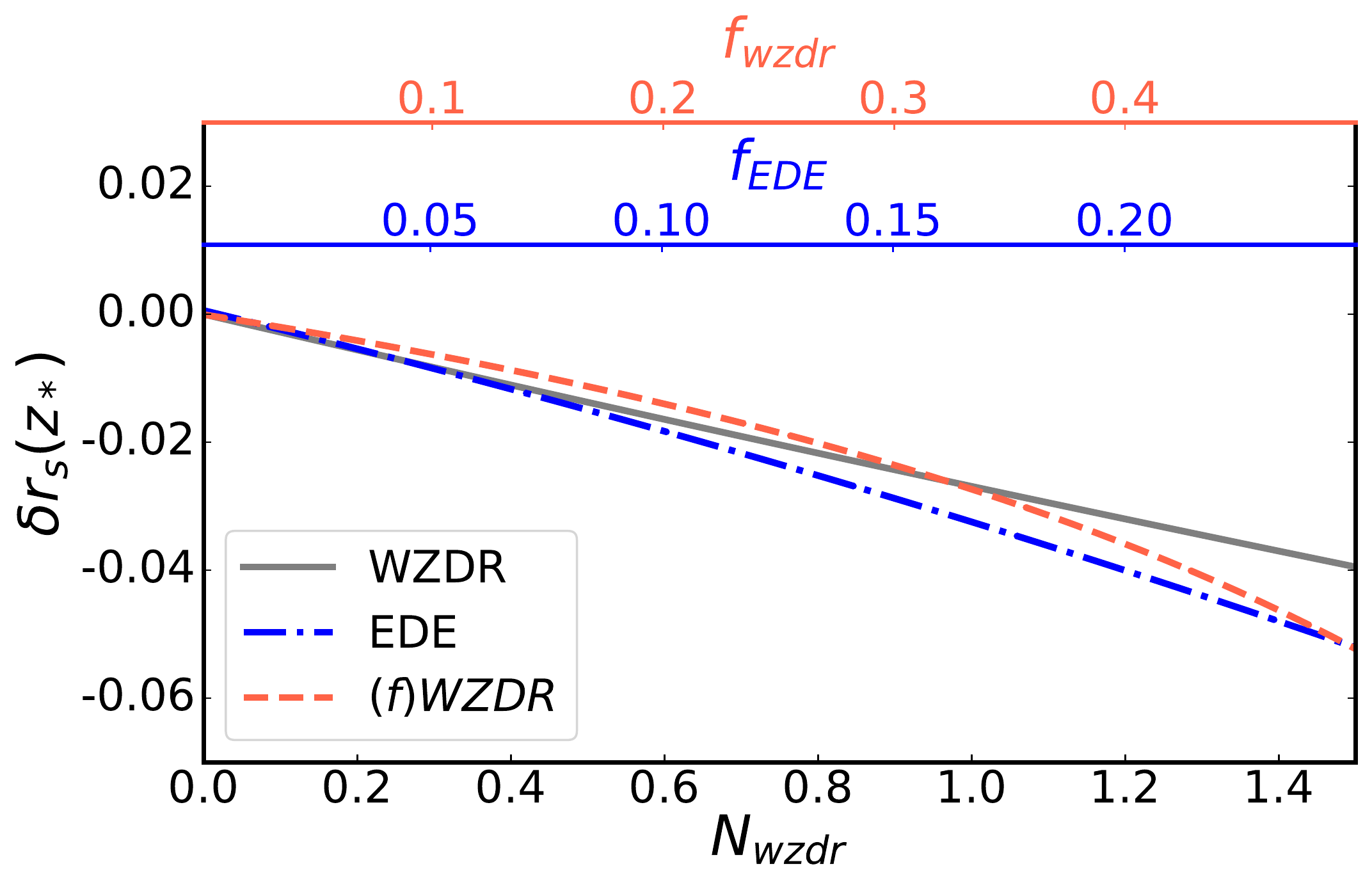}
	\includegraphics[scale=0.38]{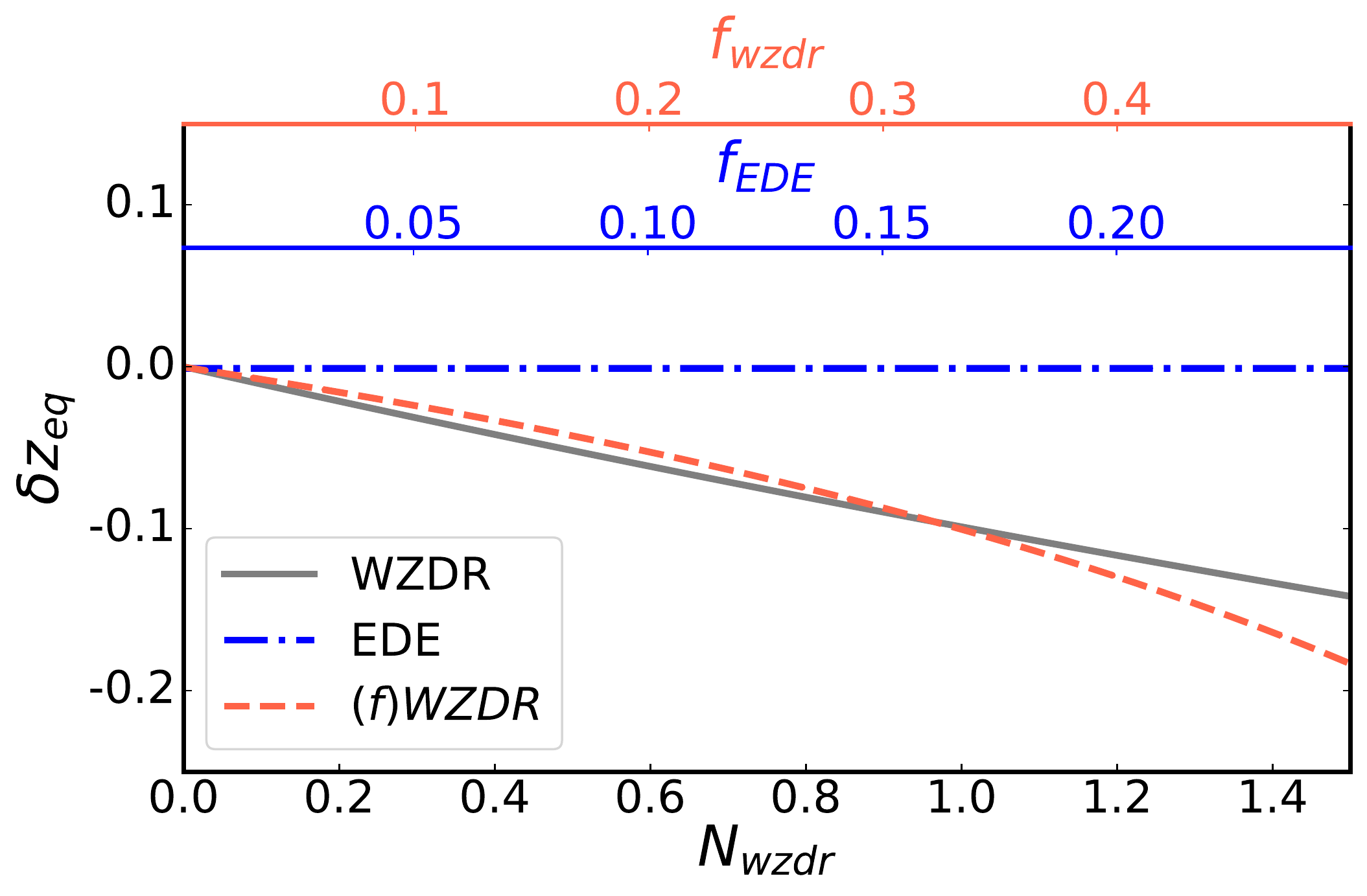}
	\caption{The relative difference between the best-fit parameters ($k_{\rm{eq}}, r_s, z_{\rm{eq}}$) calculated in fiducial $\Lambda$CDM model and its extensions (EDE, WZDR).}\label{fig:deviation}
\end{figure}

\section{Models}\label{sec:models}
In this section, we present the basic features of the WZDR model \cite{Wess:1974tw} for cosmological implementations. We refer the reader to Appendix A in ref. \cite{Aloni:2021eaq} for detailed discussions.
The interacting dark radiation consists of a Weyl fermion $\psi$ and complex scalar $\phi$, which interact through a Yukawa coupling, $\lambda\psi^2\phi$. The Lagrangian reads
\begin{equation}
\mathcal{L}_{\text{WZ}} = \lambda \phi \psi^2 + \lambda^2 (\phi^*\phi)^2.
\end{equation}
This model is referred to as the  “Wess-Zumino Dark Radiation” (WZDR). 
When the temperature of the coupled fluid drops below the scalar mass $m_\phi$, the scalars deposit their entropy into the lighter (massless) $\psi$ species. According to the entropy conservation, during this process, the lighter species would be heated, while the energy density of
interacting fluid transitions as $N_{\rm{wzdr}} \rightarrow N_{\rm{wzdr,early}}$ at a redshift $z_t$, where
\begin{equation}
N_{\rm{wzdr}} = \frac{\rho_{\rm{wzdr}}}{\rho_{1\nu}}
\end{equation}
is the effective number of WZDR in the present time, with $\rho_{1\nu}(a)$ the energy density of a single neutrino in standard $\Lambda$CDM cosmology.
In \emph{Planck} best-fit WZDR model, the transition of $N_{\rm{dr}}$ happens just before the matter radiation equality, i.e, $\log_{10} z_t \approx 4.3\pm 0.2$, which implies $m_\phi$ is of order of $1\rm{eV}$ or $10\rm{eV}$ The fluid consists of both massive particle $\phi$ 
and massless particles $\psi$ during the transition.
After the transition, the massive particle $\phi$ have annihilated away so that the energy density of fluid in the present time is dominated by massless particle $\psi$, with 
\begin{equation}
\rho_{\psi} = \frac{7}{4} \frac{\pi^2}{30}T^{4}_d = \rho_{1\nu}f^4_T,\label{eq:rho_psi}
\end{equation}
where $f_T\equiv T_\nu/T_{\rm{dr}}$ is the temperature ratio between neutrinos and WZDR. 
Accordingly, The energy and pressure density of WZDR can be written as
\begin{subequations}
	\begin{align}
	\rho_{\rm{wzdr}} \equiv \rho_{\psi} + \rho_{\phi} =  \rho_{1\nu} [1 + r_g \hat{\rho}(x)] f^4_T\label{eq:rho_dr} \\
	p_{\rm{wzdr}} \equiv p_{\psi} + p_{\phi} =  p_{1\nu} [1 + r_g \hat{p}(x)] f^4_T \label{eq:p_dr} 	
	\end{align}
\end{subequations}
where we have assumed  
\begin{equation}
	\rho_\phi = r_g \rho_\psi \hat{\rho}(x),  \quad p_\phi = r_g p_\psi \hat{p}(x),	
\end{equation}
with $x \equiv m_{\phi} /T_{\rm{dr}} $, and $r_g = 8/7$ the density ratio of bosons to fermions for a single degree of freedom. Note that the dimensionless integrals $\hat{\rho}$ and $\hat{p}$ can analytically be expressed in terms of the Bessel functions of the second kind.
According to the entropy conservation in the dark fluid
\begin{equation}
S\propto a^3\frac{\rho_{\rm{dr}} + p_{\rm{dr}}}{T_{\rm{dr}}} = \frac{\rho_{\rm{dr,0}} +p_{\rm{dr,0}}}{T_{\rm{dr,0}}},\label{eq:entropy_conserv} 
\end{equation}
with $\rho_{\rm{dr,0}} = 3\:p_{\rm{dr,0}} = N_{\rm{dr, 0}} \rho_{1\nu, 0}$. Combing Eq. (\ref{eq:rho_psi}), Eq. (\ref{eq:rho_dr}, \ref{eq:p_dr}) and Eq. (\ref{eq:entropy_conserv}) one can solve $x(a)$ numerically, while $f_T$ can be expressed in terms of $x(a)$, i.e.,
can be expressed as  
\begin{equation}
\begin{aligned}
f_T & = f_{T,0} \frac{a}{xa_t} = f_{T,0} \left( 1 + \frac{r_g}{4}[3\hat{\rho}(x) + \hat{p}(x)] \right)^{-1/3}
\end{aligned}
\end{equation}
where $f_{T,0} = N_{\rm{dr, 0}}^{\frac{1}{4}}$, and $a_t \equiv T_{dr,0}/m_\phi$ is the transition scale factor.

In this work, we assume one massive neutrino species with a mass of $0.06\rm{eV}$ and $N_{\rm{ur}}$ to be the effective number of ultra-relativistic (massless neutrino) species.
We use the fractional contribution of WZDR to total ultra-relativistic species,
\begin{equation}
	f_{\rm{wzdr}} \equiv \frac{N_{\rm{wzdr}}}{N_{\rm{wzdr}} + N_{\rm{ur}}},
\end{equation} 
to parameterize the WZDR model (denote as (f)WZDR). 
This is analogous to the EDE model, which is parameterized by 
maximal fractional contribution of the EDE component to the energy density of
the universe, $f_{\rm{EDE}} \equiv \rho_{\rm{EDE}}(z_c)/\rho_{\rm{tot}}(z_c)$ and the critical redshift $z_c$. 
The WZDR model can also be effectively described by the effective number of additional neutrino species $N_{\rm{wzdr}}$ (equivalent to $N_{\rm{IR}}$ in Ref. \cite{Aloni:2021eaq}). See Appendix \ref{sec:bf_wzdr} for the cosmological constraints on both of the parameterizations described above\footnote{$N_{\rm{ur}}$ can either a free parameter or be fixed at given value. In (f)WZDR parameterization, we find $N_{\rm{ur}} = 2.00\pm 0.27$, which closely matches the value given in standard cosmology. Consequently, we have fixed $N_{\rm{ur}}$ to 2.0328 in the following analysis.}.

\section{Impact on Standard Rulers}\label{sec:impact}

\begin{figure}
	\centering
	\includegraphics[scale=0.4]{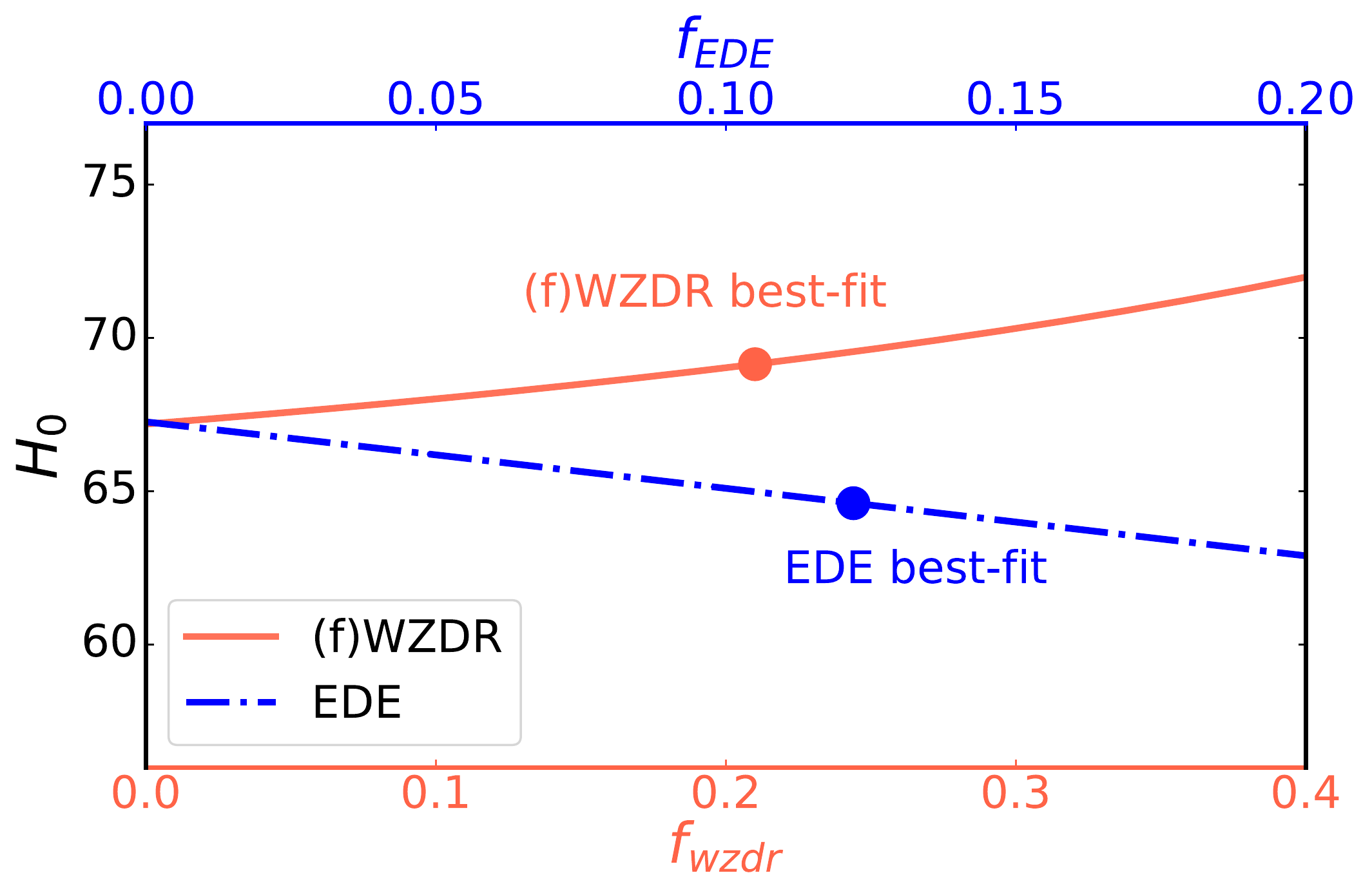}
	\caption{The derived $H_0$ in EDE and (f)WZDR scenarios with $k_{\rm{eq}}$ and $\Omega_m$ fixed at \emph{Planck} best-fit value assuming $\Lambda$CDM model.}
	\label{fig:shoot_H0}
\end{figure}

In $\Lambda$CDM model, the equality scale (in unit of $\rm{Mpc}^{-1}$) can simply be approximated as
\begin{subequations}
\begin{align}
    k_{\rm{eq}} &= (2\Omega_m H^2_0 z_{\rm{eq}})^{1/2},\label{eq:k_eq} \\
    z_{\rm{eq}} &= 2.5\times 10^{4} \Omega_{m}h^2 \Theta^{-4}_{2.7}\label{eq:z_eq}
\end{align}
\end{subequations} 
with $\Omega_m$ the matter density fraction and $\Theta_{2.7} \equiv T_{\rm{CMB}}/(2.7K)$ the CMB photon temperature.
The addition of DE component in the early time does not change the ratio between matter and radiation. Thus Eq. (\ref{eq:z_eq}) still holds in the EDE model (see the middle panel of Fig.\ref{fig:deviation}). However, Eq. (\ref{eq:k_eq}) is no longer valid in this scenario, because the the extra energy component increases the expansion rate $H(z_{\rm{eq}})$ while reduces the comoving hobble horizon at equality scale ($k^{-1}_{\rm{eq}}$). On the contrary, Eq. (\ref{eq:k_eq}) is valid in WZDR scenario whereas Eq. (\ref{eq:z_eq}) is not. The additional radiation component $N_{\rm{wzdr}}$ shifts $z_{\rm{eq}}$ to smaller value, and $k_{\rm{eq}}$ is reduced as a consequence. As is shown in Fig. \ref{fig:deviation}, the addition of EDE and WZDR shifts the standard ruler $k_{\rm{eq}}$ differently. While the sound horizon $r_s(z_*)$ is reduced in both scenarios, which increases the inferred value of $H_0$ in $r_s$-based analysis.

$H_0$ is constrained by measuring the angular scale of the cosmological horizon at matter-radiation-equality, i.e. $k_{\rm{eq}}/D_A(z) \propto k_{\rm{eq}}/h$ \cite{Philcox:2022sgj}, which is equivalent to the value of $k_{\rm{eq}}$ in unit of $h\:\rm{Mpc}^{-1}$ (If not stated otherwise, $k_{\rm{eq}}$ discussed in this paper is in unit of $h\:\rm{Mpc}^{-1}$).
As can be seen in  Eq. (\ref{eq:k_eq}, \ref{eq:z_eq}), given $k_{\rm{eq}}$ and a probe of $\Omega_m$, one can solve for the Hubble constant\cite{Philcox:2020xbv}. We have discussed how different models shift standard ruler $k_{\rm{eq}}$ compared with the fiducial $\Lambda$CDM model. Next, we fix $k_{\rm{eq}}$ and $\Omega_m$ to the value inferred in \emph{Planck} best-fit $\Lambda$CDM model\footnote{The baseline $\Lambda$CDM cosmology $\{h =0.6821,\omega_c = 0.1177,\omega_b = 0.02253\}$. We refer the readers to Section \uppercase\expandafter{\romannumeral3} A in Ref. \cite{Farren:2021grl} for detailed discussions.}
,i.e., $k_{\rm{eq}} = 1.545 \:h \rm{Mpc}^{-1}, \Omega_m = 0.311$, to see the best-fit of $H_0$ in these scenarios by ``shooting'' the targeted $k_{\rm{eq}}$ value.
We show in Fig. \ref{fig:shoot_H0} the best-fit $H_0$ value as a function of the fractional density of EDE and WZDR.
As can be seen, the increase of $k_{\rm{eq}}$ by EDE component can be compensated by a lower inferred value of $H_0$, while the decrease of $k_{\rm{eq}}$ is compensated by a higher inferred value of $H_0$. 
We will further verify this point by adopting a mock prior on $k_{\rm{eq}}$ in MCMC analysis (See Section \ref{sec:keq_rs}).

\begin{table*}
	\renewcommand{\arraystretch}{1.4}
	\centering
	\begin{tabular}{|lcc|cc|}
		\hline\hline Dataset &\multicolumn{2}{c}{FS without $r_s$ }  &\multicolumn{2}{c|}{$k_{\rm{eq}}$ prior }  \\ \hline
		Model &~~~ (f)WZDR  &EDE~~  &  (f)WZDR & EDE\\ \hline \hline
		$H_0              $   & $69.7\pm 1.1 $         & $70.6\pm 1.0\quad                 $ 
		    &      $70.3\pm 2.4$                & $65.8^{+2.0}_{-2.3}                        $ \\
		$10^{2} k_{\rm{eq}}\quad$    &  $1.517^{+0.055}_{-0.050}        $ & $1.673^{+0.048}_{-0.055}\quad$    
		&  $\quad 1.545\pm 0.020  $ & $1.545\pm 0.021$ \\
			$S_8       $ & $0.809\pm 0.027$ & $0.820^{+0.025}_{-0.029}   \quad$ 
			& $0.825\pm 0.035$ & $0.725^{+0.026}_{-0.030} $  \\
			$\Omega{}_{m }  $ & $0.310^{+0.011}_{-0.0094}           $ &$0.3129\pm 0.0088            \quad$& $0.310^{+0.011}_{-0.0094}           $&$0.3129\pm 0.0088            $\\
			$\chi^2_{\rm{min}} $ & $1072.06 $ & $1073.42$ & $ 0.06$ & $0.08$ \\
		\hline
	\end{tabular}
	\caption{The the mean $\pm1\sigma$ constraints on the derived parameters in EDE and $f$WZDR model, as inferred from base datasets $\mathcal{B}$ + FS power spectrum with $r_s$ marginalized over (left panel), and  $\mathcal{B'}$ + $k_{\rm{eq}}$ prior derived from \emph{Planck} best-fit. Upper and lower bounds correspond to the $68\%$ C.L. interval. }
	\label{tab:FS_constraints}
\end{table*}
\section{datasets and numerical method}\label{sec:datasets}
Following in Ref. \cite{Philcox:2022sgj}, we consider the following combination as our base datasets (denote as $\mathcal{B}$), which includes:
\begin{itemize}
    \item Supernovae (SNe): The derived $k_{\rm{eq}}/h$ (in unit of $\rm{Mpc}^{-1}$) is degenerate with the matter density, i.e., $k_{\rm{eq}}/h \propto \Omega_m^{1/2}h$ (see Eq. \ref{eq:k_eq}, and Eq. \ref{eq:z_eq}), thus, the additional priors on $\Omega_m$ can improve the constraints on $H_0$ from the equality scale. We consider the Gaussian prior $\Omega_m = 0.338\pm 0.018$ from the most recent Pantheon analysis \cite{Aboubrahim:2022gjb}.
    \item Big Bang Nucleosynthesis (BBN): 
    Including the $\omega_b$ prior from BBN breaks the $H_0-\omega_b$ degeneracy and will strengthen the sources of $H_0$ information from the equality scale. Consequently, a prior on the physical baryon density $\omega_b = 0.02268 \pm 0.00036$ is imposed to maximize the extracted information, following [28].
    \item  Spectral Index and Primordial Amplitude: Following from previous works [41], we add a flattened Gaussian prior on $A_s$ and $n_s$ centred at \emph{Planck} best-fit \cite{Planck:2018vyg} i.e., $n_s = 0.965\pm0.02$, and $A_s = (2.11\pm 0.108)\times 10^{-9}$.
\end{itemize}
The base datasets $\mathcal{B}$ can optionally be combined with:
\begin{itemize}
\item Galaxy Power Spectra: We adopt the method described in Ref. \cite{Farren:2021grl} to marginalized over $r_s$ information in galaxy power spectrum.
This method can be applied to the current version of the BOSS DR12 galaxy survey dataset \cite{BOSS:2016hvq}, which is part of SDSS-III \cite{SDSS-III:2015hof,BOSS:2012dmf}.
This data split across two redshift bins (at z = 0.38, 0.61) in each of the Northern and Southern galactic caps (NGC and SGC). Following from previous work, we utilize the CLASS-PT implementation \cite{Chudaykin:2020aoj} with publicly available likelihoods\footnote{Available at \url{https://github.com/oliverphilcox/full_shape_likelihoods}.}, in which the power spectrum are modelled with the Effective Field Theory of Large Scale Structure \cite{Ivanov:2019pdj} including one-loop perturbation theory and etc.
\item Baryon Acoustic Oscillation: measurements of the BAO signal at $z = 0.38, 0.51$
and $0.61$ from BOSS DR12 galaxy power spectrum \cite{Alam:2016hwk}.
\item $S_8$ prior: We adopt the prior $S_8 = 0.800\pm0.029$ from KiDS-450+GAMA \cite{vanUitert:2017ieu}, similar results can be found in the the measurement from the first-year data of HSC SSP20, for which $S_8 = 0.804\pm0.032$ \cite{Hamana:2019etx} .           
\end{itemize} 
We are interested in how the extra radiation (DE) component affects the Hubble constant inferred from the standard ruler $k_{\rm{eq}}$. Hence, throughout the analysis we fix the fractional density of additional components to best-fit value obtained in cosmological analysis (see Appendix \ref{sec:bf_wzdr} for a detailed discussion of the best-fit (f)WZDR model), i.e., $f_{\rm{wzdr}} = 0.21$ and $f_{\rm{EDE}} = 0.122$\footnote{We adopt the best-fit values of EDE model from Table I of Ref. \cite{Smith:2019ihp}}. We sample over the following set of cosmological parameters:
\begin{equation}
    \{H_0, \omega_{\rm b}, \omega_{\rm cdm}, \log 10^{10}A_s, n_s, \sum m_\nu\}.
\end{equation}
The optical depth of reionization, $\tau_{\rm reio}$, is fixed to $0.055$.
We implement the EDE and the WZDR scenarios as modifications to
the publicly available Einstein-Boltzmann code \href{http://class-code.net/}{\texttt{CLASS}} \cite{Lesgourgues:2011re,Blas:2011rf} package.
The non-linear matter power spectrum required by redshift-space distortion (RSD) likelihoods are computed using the ``HMcode'' \cite{Mead:2015yca,Mead:2016ybv,Mead:2020vgs} implemented in \texttt{CLASS}.
The MCMC analyses are performed using the publicly available code MontePython \cite{Brinckmann:2018cvx}.

\section{numerical results}\label{sec:results}

\begin{figure*}
	\centering
	\subfigure[ $\:$FS-$r_s$ ]{
	\includegraphics[scale=0.4]{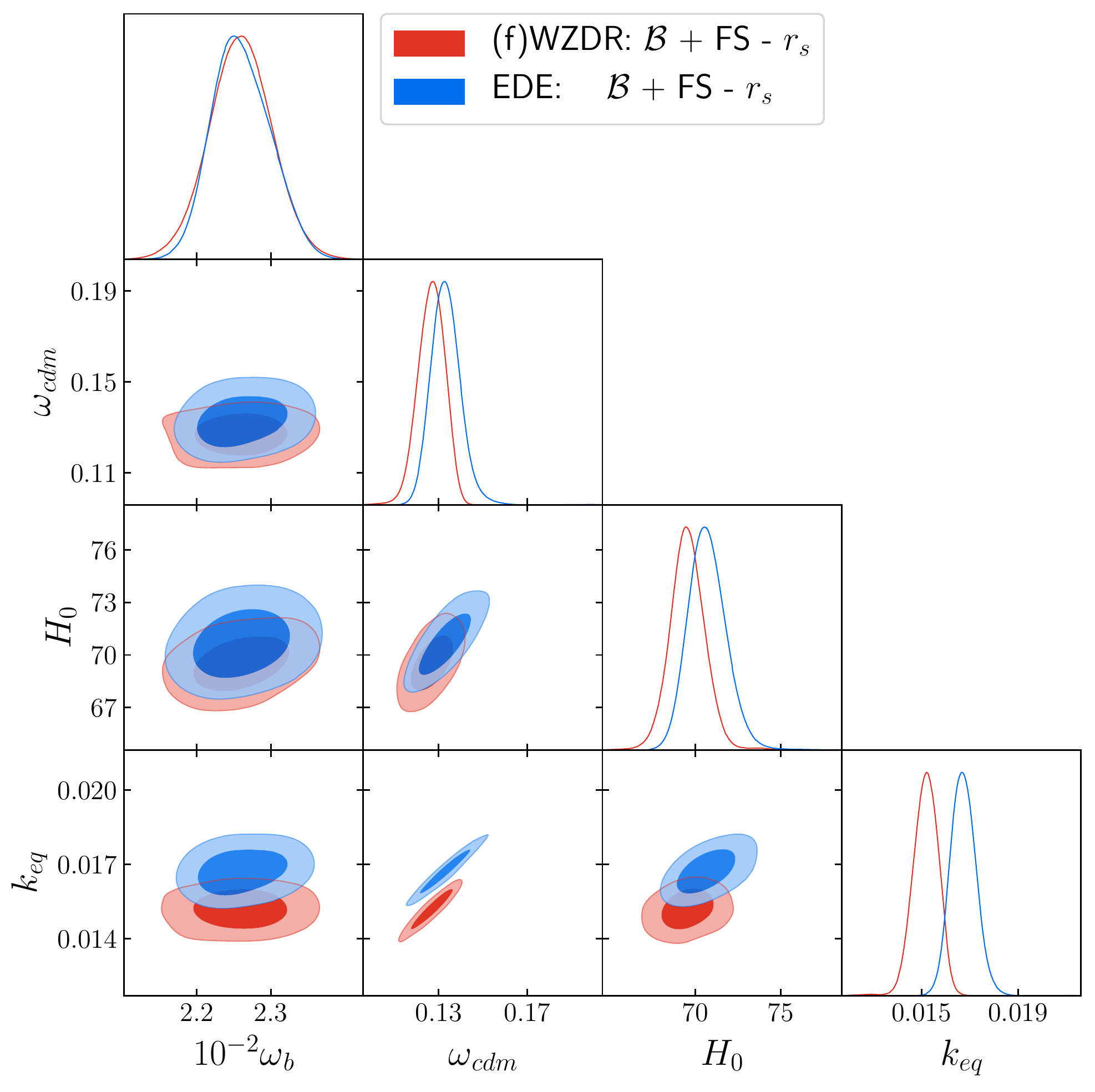}\label{fig:FS_plot}
	}
	\subfigure[ $\:k_{\rm{eq}}$ ]{
	\includegraphics[scale=0.4]{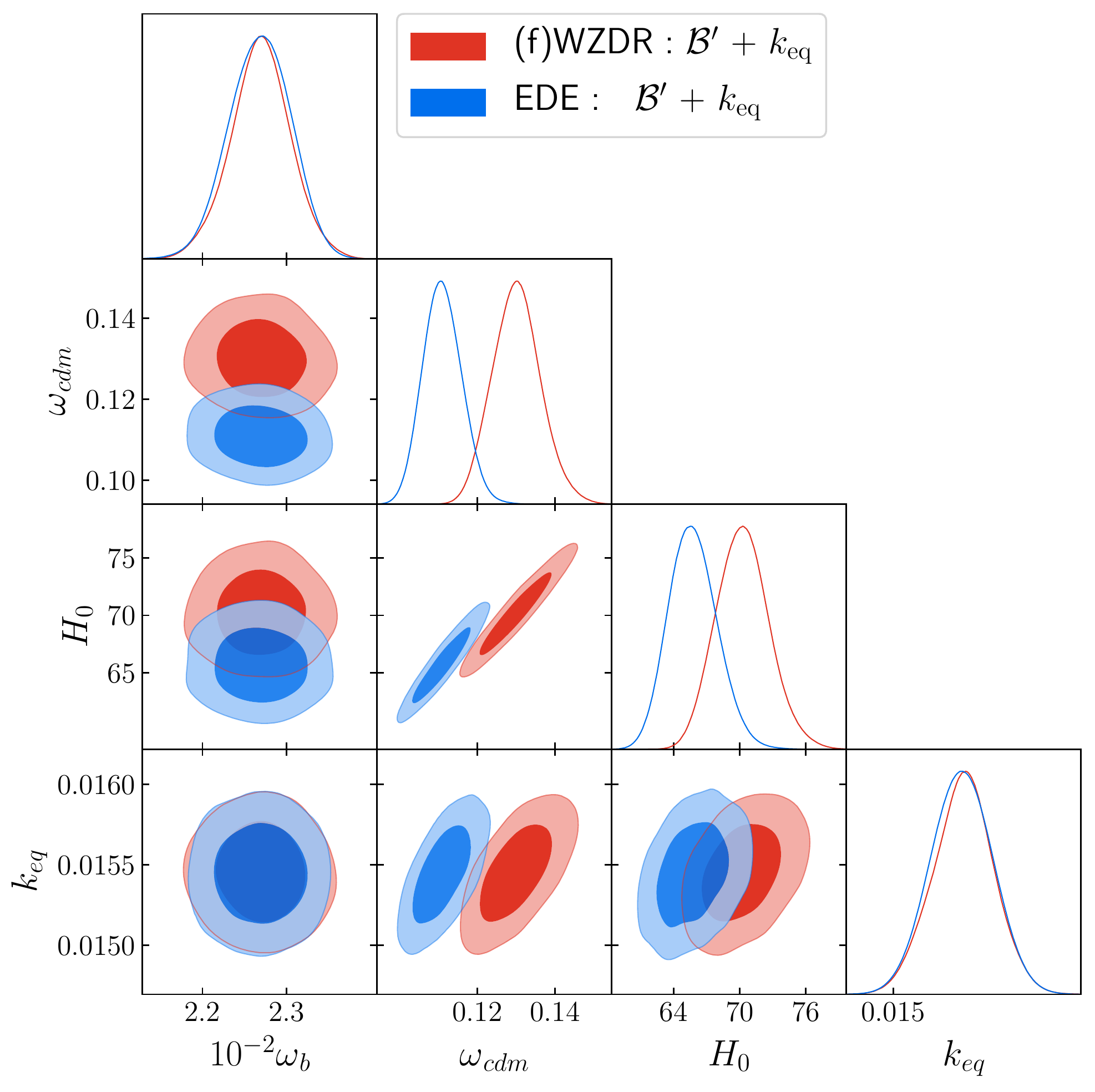}\label{fig:keq_plot}
	}
	\caption{Cosmological parameters constraints for EDE model (blue contour) and (f)WZDR model (red contour) from the dataset: $\mathcal{B}$ + FS power spectrum with $r_s$ marginalized over (denoting as FS - $r_s$ in the left panel), and fitting to $\mathcal{B'}$ + $k_{\rm{eq}}$ prior derived from \emph{Planck} best-fit $\Lambda$CDM model (right panel).}
	\label{fig:FS_contour}
\end{figure*}
We show in Fig. \ref{fig:FS_plot} and Tab. \ref{tab:FS_constraints} the $r_s$-independent constraints from FS galaxy power spectra, supplemented by the base datasets $\mathcal{B}$ which includes priors on $\Omega_m$ from \textsc{Pantheon+} and $\omega_b$ from BBN. 

\begin{figure*}
	\centering
	\includegraphics[scale=0.5]{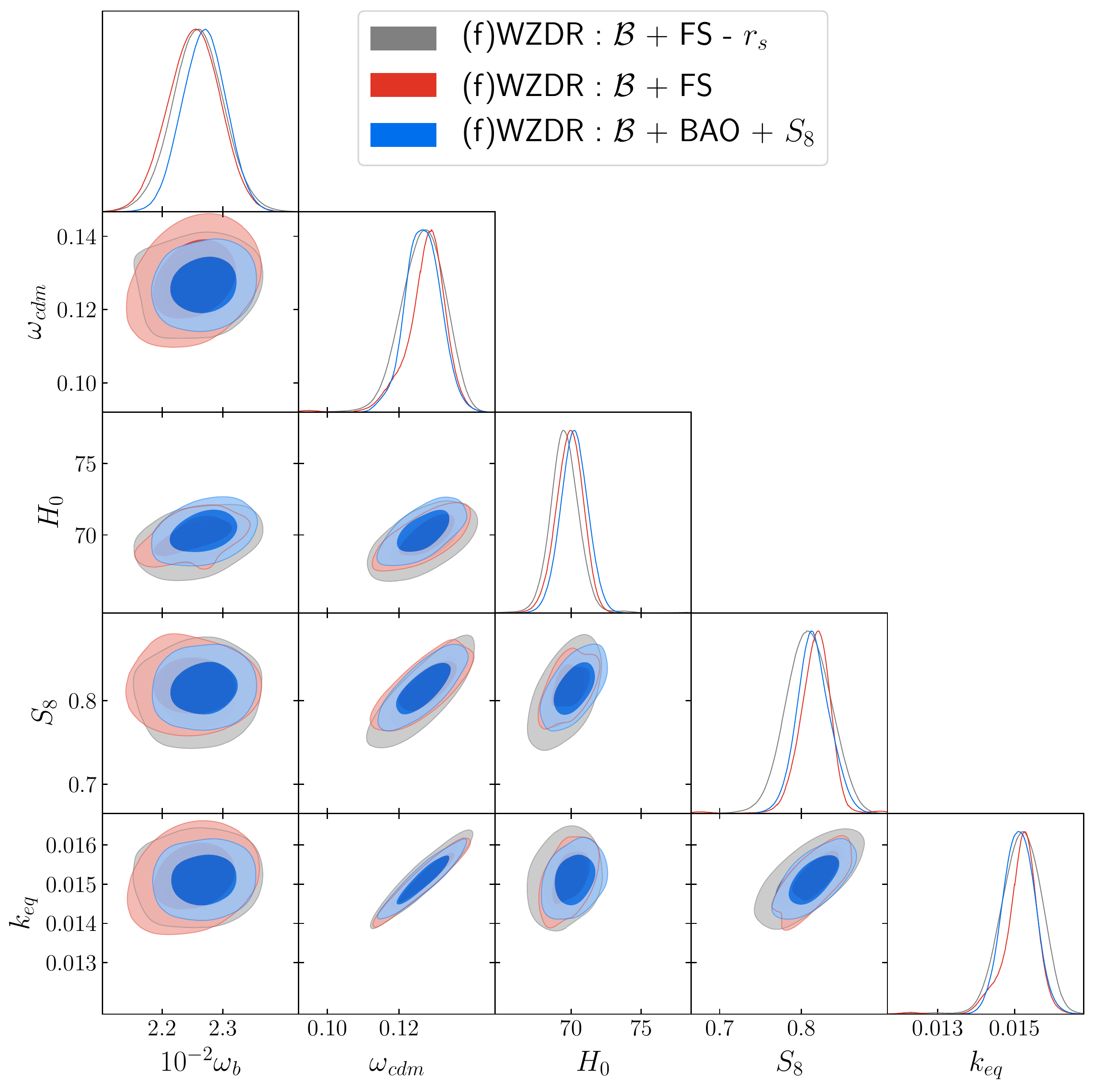}
	\caption{Constraints on cosmological parameters in the WZDR scenario from 
	the base datasets $\mathcal{B}$ combined with three choices of likelihoods: (1), the FS likelihoods with $r_s$ marginalized over, (2), the FS likelihoods including the BAO information, and (3), the BAO likelihoods combined with $S_8$ prior derived from lensing experiments.}
	\label{fig:WZDR_contour}
\end{figure*}

\begin{figure*}
	\centering
	\includegraphics[scale=0.5]{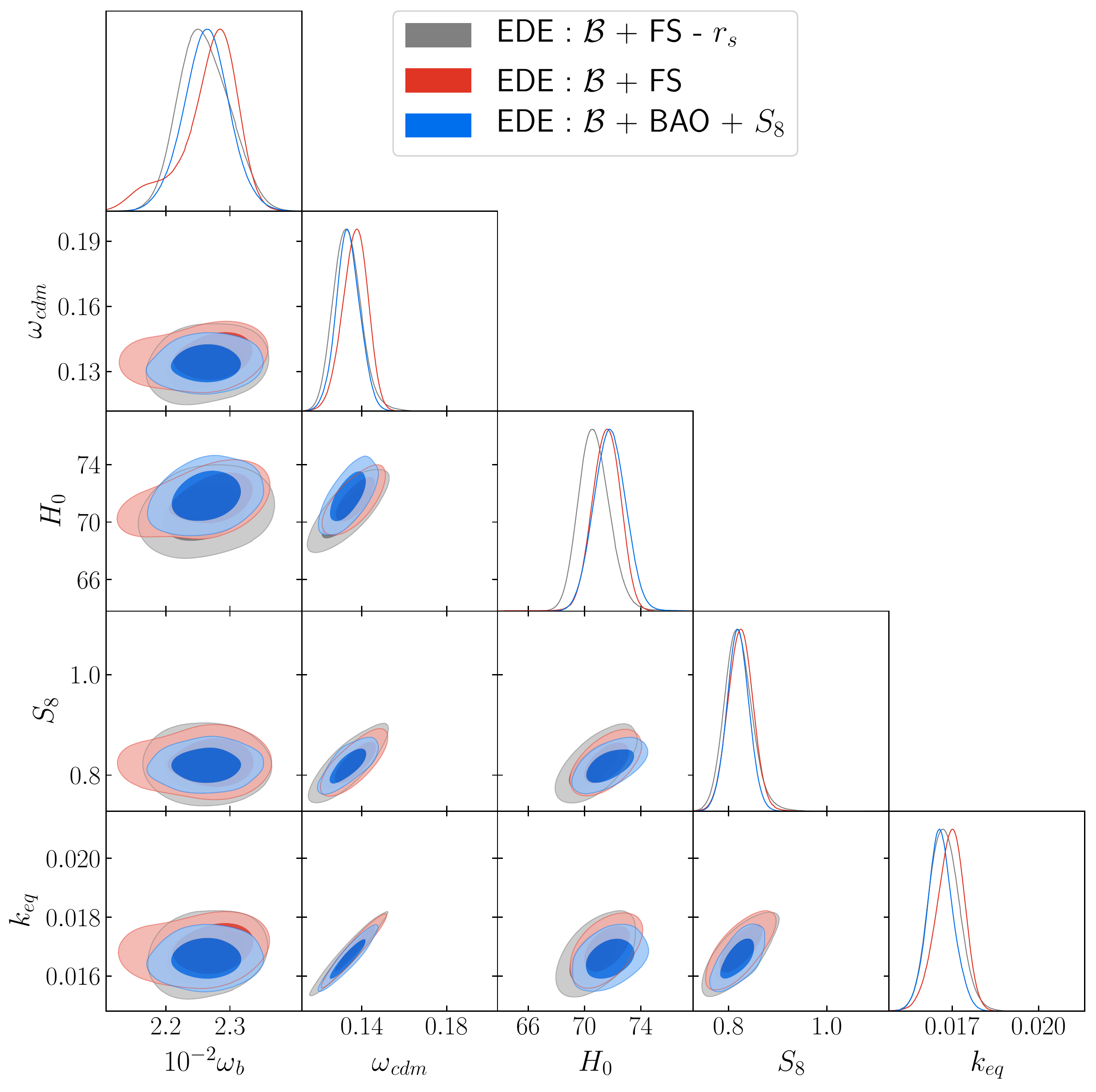}
	\caption{Constraints on cosmological parameters in the EDE scenario from three choices of likelihoods discussed above.}
	\label{fig:EDE_contour}
\end{figure*}

For comparison between constraints with and without $r_s$ information, we show in Table \ref{tab:main_constraints} the parameter constraints obtained from the base datasets $\mathcal{B}$ combined with three choices of likelihoods: (1), the FS likelihoods with $r_s$ marginalized over, (2), the FS likelihoods including the BAO information, and (3), the BAO likelihoods combined with $S_8$ prior derived from lensing experiments. The triangular plot with the 1D posterior distributions and the 2D contour plots for the constrained parameters are shown in Fig. \ref{fig:WZDR_contour} and Fig. \ref{fig:EDE_contour}.

\subsection{Derived equality scale }
In the discussion of Sec. \ref{sec:impact}, we have shown that the addition of the WZDR reduces $k_{\rm{eq}}$ while the opposite is true for the EDE component. 
As is shown in Table \ref{tab:main_constraints}, the $1\sigma$ constraint of $k_{\rm{eq}}$ from $r_s$-based analysis (BAO+$S_8$) in (f)WZDR model, $k_{\rm{eq}} = 1.512 \pm 0.041$, is $2.4\sigma$ lower than that in EDE model, $k_{\rm{eq}} = 1.659 \pm 0.045$.
Still, there is a $2.2\sigma$ tension between the EDE and (f)WZDR best-fit $k_{\rm{eq}}$ when fitting with the $k_{\rm{eq}}$-based analysis ($r_s$-marginalized FS power spectra). While in each model, $k_{\rm{eq}}$ derived from the BAO+$S_8$ prior is close to that from the $r_s$-marginalized FS power spectrum.
These results suggest that $k_{\rm{eq}}$ can not yet be tightly (model independently) constrained by the Equality-based measurements.   
Currently, the information extracted from the $r_s$-marginalized FS power spectrum is dominated by information concerning structure growth ($f\sigma_8$, $S_8$) rather than the angular scale of the cosmological horizon at matter-radiation-equality. Otherwise, the derived $k_{\rm{eq}}$ in $k_{\rm{eq}}$-based analysis from different models should be more consistent.

\subsection{$k_{\rm{eq}}$- and $r_s$-based inferrance of $H_0$}\label{sec:keq_rs}
In WZDR model, we find a minor peak shift of $Delta H_0 = 0.2$ between the fit of FS power spectrum with and without $r_s$ information, and a peak shift of $Delta H_0 = 0.4$ from the $k_{\rm{eq}}$- to $r_s$-based analyses (see Table \ref{tab:main_constraints}). While in EDE model we have $Delta H_0 = 1.0$ between FS and FS - $r_s$, and $Delta H_0 = 1.2$ between $k_{\rm{eq}}$- and $r_s$-based analyses.
Since the addition of radiation component shifts the two standard rulers more coherently (see the discussion in Sec. \ref{sec:impact}), the best-fit $H_0$ from $r_s$- and $k_{\rm{eq}}-$based analyses in the WZDR model are more consistent than that in the EDE model.
However, the above results are inconclusive since the Hubble constant measured by the two rulers is consistent (within $2\sigma$) for both EDE and WZDR models. 
As is discussed above, $k_{\rm{eq}}$-based datasets are dominated by the information $f\sigma_8$ ($S_8$) rather than the angular scale of the cosmological horizon at matter-radiation-equality, which makes it difficult for us to detect the inconsistency.

To see how the standard ruler $k_{\rm{eq}}$ affects the infferance of $H_0$, we replace the $r_s$-marginalized FS power spectra with a mock prior: $k_{\rm{eq}} =1.545\pm 0.020\times 10^{-2}h \rm{Mpc}^{-1}$ derived from \emph{Planck} best-fit $\Lambda$CDM model (Accordingly, the $\Omega_m$ prior in the supplement datasets $\mathcal{B}$ are substituted by the emph{Planck} best-fit value, $\Omega_m$, denoting as  $\mathcal{B'}$). 
The results can be found in the right panel of Tab. \ref{tab:FS_constraints}.
As is discussed in Sec. \ref{sec:impact}, to fit with the standard rulers fixed by the fiducial $\Lambda$CDM model, the increase of $k_{\rm{eq}}$ should be compensated by a lower $H_0$. Accordingly, its decrease should be compensated by a higher $H_0$. In this case, the inferred value of $H_0$ in the WZDR model is significantly higher than in the EDE model. Note that we have also adopted a tight prior on $\Omega_m$, so higher $H_0$ will result in higher $\omega_c = \Omega_m h^2 - \omega_b$ and higher $S_8$. The inferred $S_8$ in the EDE model is much lesser than that of the WZDR model.

\begin{table*}
	\renewcommand{\arraystretch}{1.2}
	\begin{center}
		\begin{tabular}{|llccccccc|}			
			\hline
			\hline			
			Model\hspace{2pt} &Dataset & $H_0$ & $10^{2}k_{\rm{eq}}$ & $S_8$ & $\Omega_m$ & {\boldmath$\ln10^{10}A_{s }$} &{\boldmath$n_{s}         $}&$\chi^2_{\rm{min}} $ \\
			\hline
			\hline
			
			(f)WZDR$\:$ 
			& $\mathcal{B}$+FS - $r_s$&$69.7\pm 1.1$ &$1.517^{+0.055}_{-0.050}$& $0.809\pm 0.027$ & $0.310^{+0.011}_{-0.0094}$ &  $3.024^{+0.039}_{-0.031}$  &$0.973\pm 0.019            $&1072.06 \\
			& $\mathcal{B}$+FS &  $69.90^{+0.90}_{-0.79} $ &$1.515^{+0.047}_{-0.029}        $& $0.817^{+0.020}_{-0.016} $ & $0.308^{+0.011}_{-0.0042} $ & $3.036^{+0.023}_{-0.034} $ &$0.971^{+0.021}_{-0.026}$&1072.26\\
			&$\mathcal{B}$+BAO+$S_8$ & $70.13\pm 0.96   $ &$1.512\pm 0.041        $& $0.813^{+0.019}_{-0.021}$ & $0.3042\pm 0.0082$ &  $3.037\pm 0.033 $ &$0.963^{+0.016}_{-0.022}   $&7.78\\
			\hline
			\hline
			EDE 
			& $\mathcal{B}$+FS - $r_s$& $70.6\pm 1.0$ &$1.673^{+0.048}_{-0.055}$& $0.820^{+0.025}_{-0.029}$ & $0.3129\pm 0.0088$ & $3.030\pm 0.036$ & $0.975\pm 0.020   $ &1073.42 \\
			& $\mathcal{B}$+FS &  $71.6\pm 0.9 $ &$1.692^{+0.053}_{-0.044}$ & $0.820\pm 0.023  $ & $0.3134\pm 0.0085$ & $3.000^{+0.035}_{-0.042}$ &$0.969^{+0.018}_{-0.020}      $&1073.58\\
			&$\mathcal{B}$+BAO+$S_8$ & $71.8\pm 1.1   $ &$1.659\pm 0.045        $& $0.819\pm 0.022$ & $0.3043^{+0.0076}_{-0.0091}$ &  $3.032\pm 0.040  $ &$0.964^{+0.015}_{-0.020}   $&8.18\\
			\hline
			
		\end{tabular}
	\end{center}
	\caption{The mean and $1$-$\sigma$ constraints on cosmological parameters in the (f)WZDR model (top panel) from the base datasets $\mathcal{B}$ combined with three choices of likelihoods: (1), the FS likelihoods with $r_s$ marginalized over (denoting as FS - $r_s$), (2), the FS likelihoods including the BAO information, and (3), the BAO likelihoods combined with $S_8$ prior derived from lensing experiments. Upper and lower bounds correspond to the $68\%$ C.L. interval.}
	\label{tab:main_constraints}
\end{table*}

\section{CONCLUSIONS}\label{sec:conclusions}
The Hubble horizon at matter-radiation equality, $k^{-1}_{\rm{eq}}$, provides valuable test of new physics 
around (prior to) $z_{\rm{eq}}$, while the commonly used standard ruler, $r_s$, is sensitive to physics prior to the last scattering.
These two rulers provide an interesting consistency check for the $\Lambda$CDM model and its extensions,
i.e., the inferred value of $H_0$ from both the rulers should be consistent.
We find that the addition of EDE component which peaked around the equality scale ($\log_{10} z_c \sim 3.5$) increases equality scale, $k_{\rm{eq}}$. 
On the contray, the addition of radiational component before the equality reduces $z_{\rm{eq}}$ and $k_{\rm{eq}}$.
We show in Fig. \ref{fig:shoot_H0} that the increase of $k_{\rm{eq}}$ by EDE component can be compensated by an upward shift of $H_0$, while the decrease of $k_{\rm{eq}}$ is compensated by a higher inferred value of $H_0$.  
The two standard rulers should be shifted coherently to relieve the $H_0$ discrepancy between the early and late universe.

Our numerical results shows that, the best-fit $H_0$ values obtained from $r_s$- and $k_{\rm{eq}}-$based analyses are consistent within $1 \sigma$ in (f)WZDR model ($\Delta H_0 = 0.4$). In the same analysis assuming the EDE model, we observed a peck-shift of $\Delta H_0 = 1.2$.
There is a significant tension between the derived $k_{\rm{eq}}$ in EDE and the (f)WZDR model. Meanwhile, in each model, $k_{\rm{eq}}$ derived from the BAO+$S_8$ prior is close to that from the $r_s$-marginalized FS power spectrum.  
These results suggest that, currently, the $r_s$ marginalized FS power can not tightly contain the standard ruler $k_{\rm{eq}}$. The extracted signal consists mainly of the information concerning the structure growth ($f\sigma_8$, $S_8$) amplitude rather than the angular scale of the cosmological horizon at matter-radiation-equality.

We expect more information on $k_{\rm{eq}}$ to be extracted from next generation of galaxy survey, e.g., Euclid\href{https://www.euclid-ec.org}, SKA\href{https://www.skatelescope.org}. As a forecast analysis, we adopt a mock prior derived from \emph{Planck} best-fit $\Lambda$CDM model to see how the standard ruler $k_{\rm{eq}}$ affects the inference of $H_0$. 
In WZDR model, we find $\Delta H_0 = 1.1$ compared with the best-fit $H_0$ in baseline $\Lambda$CDM model, while in the EDE model the pair difference is $\Delta H_0 = - 2.4$.
This result further suggests that the two standard rulers should be reduced simultaneously to relieve the tension with the local distance ladder. 

\begin{acknowledgments}
	This work is supported in part by National Natural Science Foundation of China under Grant No.12075042, Grant No.11675032 (People's Republic of China), Grant No.12175095, and supported by LiaoNing Revitalization Talents Program (XLYC2007047).
\end{acknowledgments}

\appendix
\section{Best-fit WZDR}\label{sec:bf_wzdr}
\begin{figure*}
	\centering
	\includegraphics[scale=0.36]{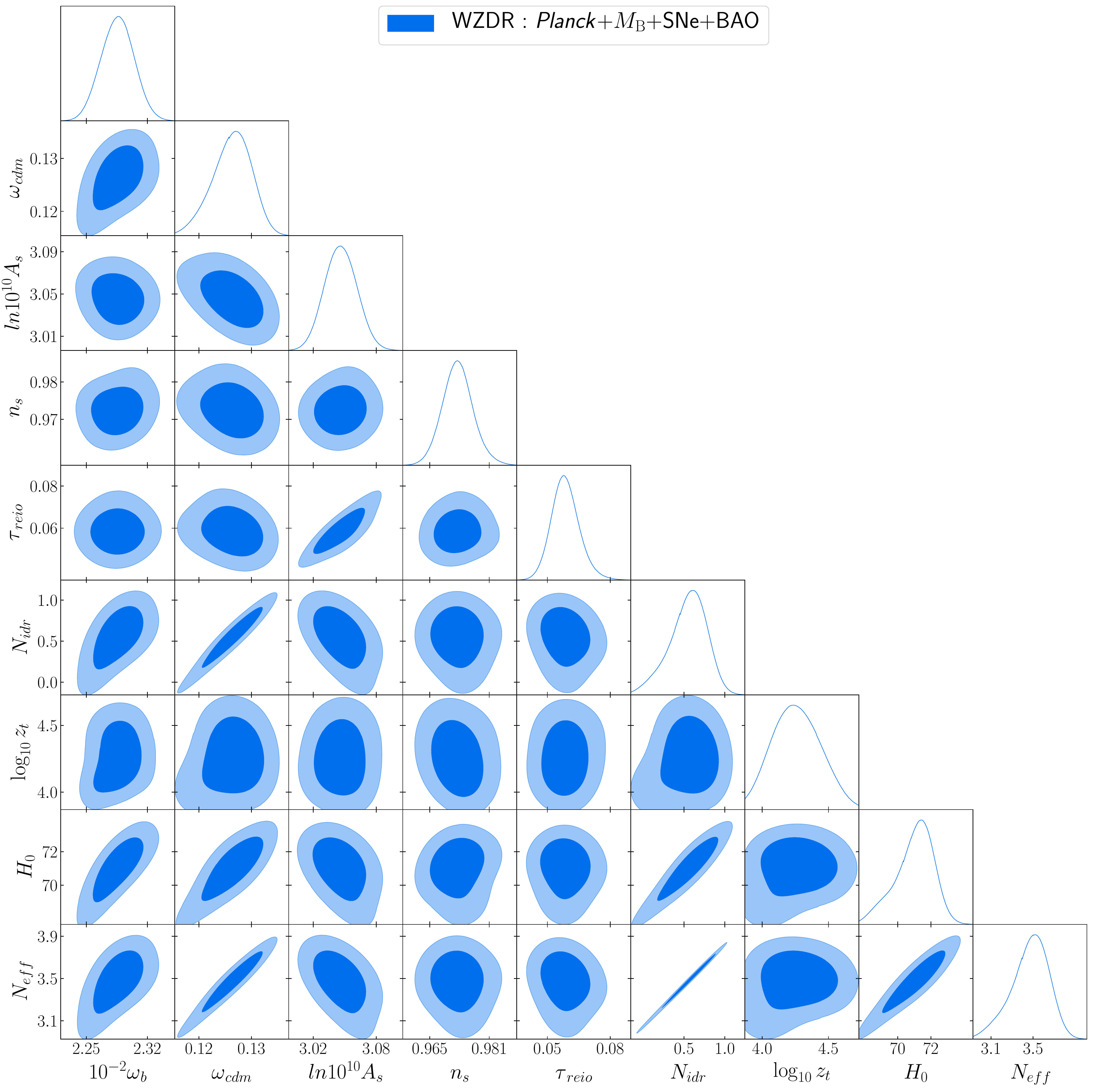}
	\caption{Cosmological parameters constraints in WZDR model (with $N_{\rm{wzdr}}$ being the only extended parameter) from the datasets discussed in Appendix \ref{sec:bf_wzdr}.}
	\label{fig:wzdr_dp}
\end{figure*}

\begin{figure*}
	\centering
	\includegraphics[scale=0.34]{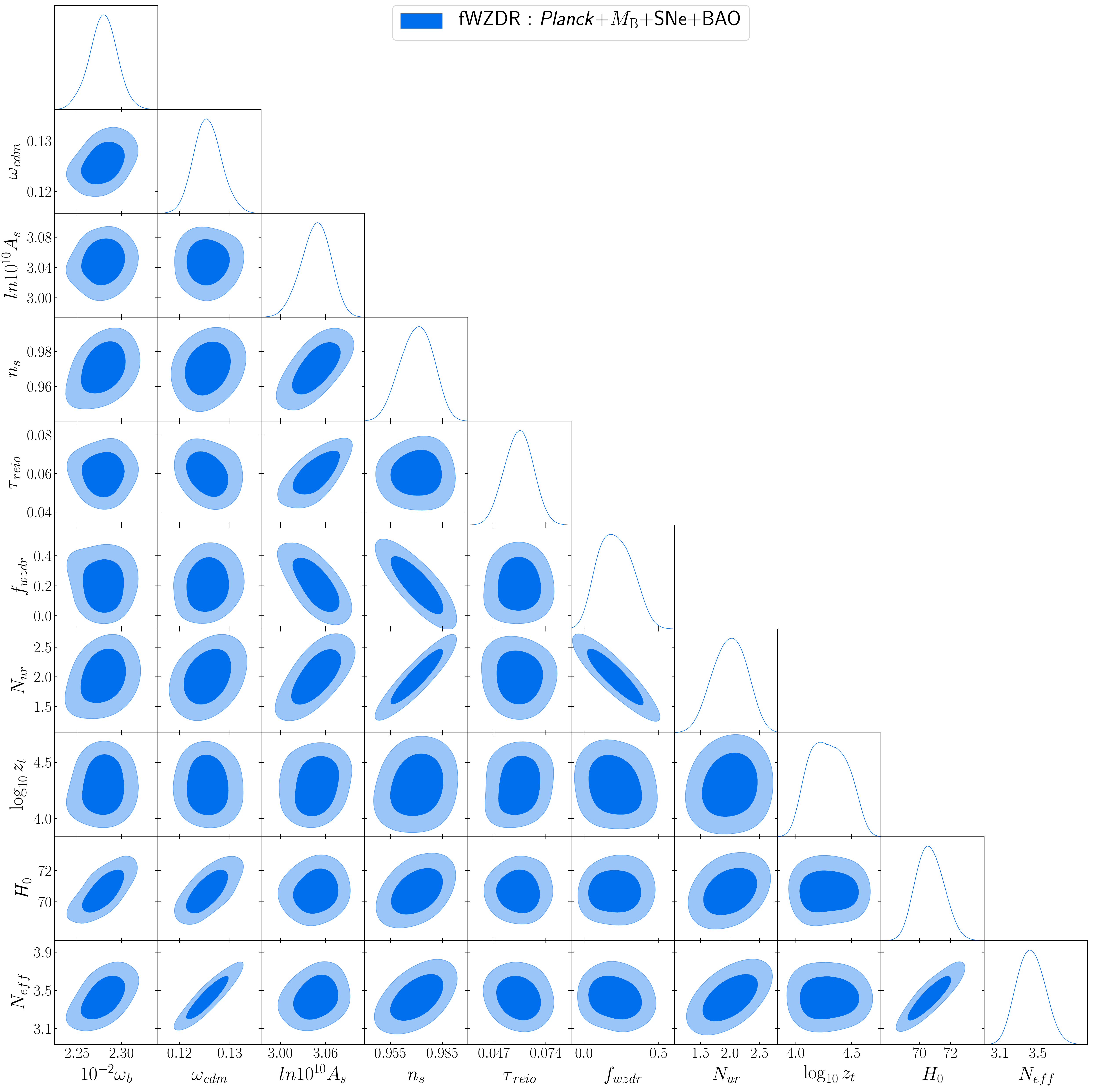}
	\caption{Cosmological parameters constraints in (f)WZDR model (with $f_{\rm{wzdr}}$ and $N_{\rm{ur}}$ being free parameters) from the datasets discussed in Appendix \ref{sec:bf_wzdr}.}
	\label{fig:wzdr_frac}
\end{figure*}

We repeat the cosmological constraints on the WZDR model with the same datasets discussed in Ref. \cite{Schoneberg:2021qvd}. Meanwhile, we adopt a new parameterization with $N_{\rm{ur}}$ being a free parameter (denote as (f)WZDR).  This datasets includes: the Planck \emph{Planck} 2018 low-$\ell$ TT+EE and high-$\ell$ TT+TE+EE temperature and polarization power spectrum \cite{Planck:2019nip,Planck:2018vyg}, as well as CMB lensing \cite{Planck:2018vyg}, the BAO measurements \cite{Alam:2016hwk, 6dF, MGS}, the PANTHEON supernovae data \cite{Scolnic:2017caz}, as well as measurements from SH0ES \cite{Riess:2021jrx} via a prior on the intrinsic magnitude of supernovae $M_b$ \cite{Camarena:2021jlr}. Constraints at $68\%$ C.L. on the cosmological parameters and $\chi^2$ statistics can be found in Table \ref{tab:wzdr_fit}. The posterior distributions and for WZDR and (f)WZDR parameterization are shown in Fig. \ref{fig:wzdr_dp} and \ref{fig:wzdr_frac}, respectively. 
\begin{table*}
	\renewcommand{\arraystretch}{1.3}
	\begin{center}
		\begin{tabular}{|l c c|}			
			\hline
			\hline			
			Parameter & WZDR & (f)WZDR   \\
			\hline
			\hline
			{\boldmath$10^{-2}\omega{}_{b }$} & $2.286\pm 0.018    $& $2.279\pm 0.016            $\\
	
			{\boldmath$\omega{}_{c}$} & $0.1263^{+0.0041}_{-0.0033}$ & $0.1256^{+0.0026}_{-0.0029}$\\
			
			{\boldmath$\ln10^{10}A_{s }$} & $3.046\pm 0.014            $& $1.04241\pm 0.00086        $\\
			
			{\boldmath$n_{s}         $} & $0.9725^{+0.0036}_{-0.0040}$ & $3.047^{+0.021}_{-0.018}   $\\
			
			{\boldmath$\tau_{\rm{reio}} $} & $0.0585^{+0.0059}_{-0.0069}$ &$0.970^{+0.010}_{-0.0092}  $\\
			
			{\boldmath$N_{\rm{idr}}       $} & $0.55^{+0.26}_{-0.19}      $ &$-$ \\

			{\boldmath$f_{\mathrm{wzdr}}          $} &$-$ &$0.21^{+0.11}_{-0.13}      $\\

			{\boldmath$N_{\mathrm{ur}}        $} &$(2.0328)$& $2.00\pm 0.27              $\\
			
			{\boldmath$log10z_{t }    $} & $4.26^{+0.18}_{-0.20}      $& $4.29^{+0.16}_{-0.19}      $\\	
			
			{\boldmath$H_0             $} & $71.0^{+1.3}_{-0.87}       $& $70.68\pm 0.89             $\\	
			
			{\boldmath$S_8         $} & $0.8185\pm 0.0059          $& $0.818^{+0.011}_{-0.010}   $\\
			
			{\boldmath$N_{\rm{eff}}           $} & $3.47^{+0.20}_{-0.15}      $& $3.43^{+0.14}_{-0.16}      $\\
			\hline

		\end{tabular}
	\end{center}
	\caption{The the mean $\pm1\sigma$ constraints on the cosmological parameters and derived parameters in WZDR and (f)WZDR model from the datasets discussed in Appendix \ref{sec:bf_wzdr}. Upper and lower bounds correspond to the $68\%$ C.L. interval. }
	\label{tab:wzdr_fit}
\end{table*}

\bibliography{wzdr}

\end{document}